 \documentclass[final,3p]{elsarticle}

\usepackage{amsmath,amssymb}
\usepackage{lipsum}
\usepackage{graphicx}
\usepackage{subfloat}
\usepackage{subfig}
\usepackage{enumerate}
\usepackage{url}
\usepackage{relsize}
\usepackage{footmisc}
\usepackage[colorlinks = true, urlcolor  = blue]{hyperref}

\journal{Physics of the Dark Universe}

\begin{document}

\begin{frontmatter}



\title{Geometrical-optics analysis of the interaction between light and gravitational waves from binaries}


\author[a,b]{Dong-Hoon Kim}
\ead{ki13130@gmail.com}
\affiliation[a]{organization={The Research Institute of Basic Science, Seoul National University},%Department and Organization
            addressline={1 Gwanak-ro}, 
            city={Seoul},
            postcode={08826}, 
            country={Republic of Korea}}
\affiliation[b]{organization={Department of Physics and Astronomy, Seoul National University},%Department and Organization
            addressline={1 Gwanak-ro}, 
            city={Seoul},
            postcode={08826}, 
            country={Republic of Korea}}

\begin{abstract}
We consider a situation in which light emitted from the neighborhood of a binary interacts with gravitational waves from the binary (e.g., a supermassive black hole binary in a quasar, 
a binary pulsar, etc.). The effect is cumulative over the long path lengths of light propagation and might be appreciable if the interaction initially takes place close to the source of gravitational waves, 
where the strain amplitude can be large. This situation can be modeled effectively using spherical gravitational waves (i.e., transverse-traceless radially propagating waves), with the strain amplitude 
varying with the distance from the source to a field point where the two wavefronts of light and gravitational waves meet each other. Our analysis employs geometrical-optics methods in curved spacetime, 
where the curvature is due to gravitational waves propagating in a flat spacetime background. We place a particular focus on the effect of gravitational Faraday rotation (or Skrotskii/Rytov effect) 
resulting from the interaction between light and gravitational waves from binaries.
\end{abstract}




\end{frontmatter}

%\tableofcontents



\section{Introduction}
\label{intro}
Astronomers use light, i.e., electromagnetic waves (EMWs), to make observations of a wide variety of astrophysical events. Among others, recently, light has been used as a major tool to detect 
gravitational waves (GWs), due to its prominent property -- interaction with other kinds of waves; e.g., the property is exploited by the detection schemes via laser interferometers -- LIGO, VIRGO, 
GEO600, KAGRA, LIGO-India, eLISA, etc. \cite{PhysRevLett.116.061102,Grote_2010,Somiya_2012,LIGO-India_2011,amaroseoane_2012} 
and via pulsar timing arrays -- EPTA, PPTA, IPTA, SKA, etc. \cite{Kramer_2013,Hobbs_2013,Manchester_2013,Dewdney_2009}. 

Two massive objects orbiting each other (e.g., a black hole (BH) binary, a neutron star (NS) binary, a BH-NS binary, etc.) generate GWs, which propagate radially in all directions, with the energy of GWs decreasing 
over the distance from the source. However, when reaching a distant observer (like ones located at those aforementioned detectors), the energy of GWs decreases considerably, and the (strain) amplitude is usually 
treated as an extremely small constant value; therefore, the effects on EMWs due to GWs are extremely small too, of the same order as the strain amplitude.

In case light is emitted from the neighborhood of a binary (e.g., a supermassive BH binary in a quasar, a binary pulsar, etc.) at the same time GWs are emitted from the binary, the EMWs interact with the GWs 
all along the propagation path towards an observer. Here the strain amplitude varies with the distance from the binary source to a field point of GWs, where the two wavefronts of EMWs and GWs cross each other. 

A number of studies have been published in the context of the interaction between EMWs and plane GWs. Among others, Halilsoy and Gurtug \cite{Halilsoy_2007} analyzed the Faraday rotation 
in the polarization vector of a linearly polarized electromagnetic shock wave upon encountering with GWs. Hacyan \cite{Hacyan_2012,Hacyan_2016} determined the influence of a GW on the elliptic polarization 
of light, deducing the rotation of the polarization angle and the corresponding Stokes parameters, and applied this effect to the detection of GWs, as a complement to the pulsar timing method. 
Cabral and Lobo \cite{Cabral_2017} obtained electromagnetic field oscillations induced by GWs, and found that these lead to the presence of longitudinal modes and dynamical polarization patterns of electromagnetic radiation. 
Park and Kim \cite{Park_2021} employed the perturbation theory of general relativity to analyze the influence of GWs on EMWs, concentrating mainly on the effects on the polarization of light, and applied their analysis 
to the observation of GWs by means of Stokes parameters. Kim and Park \cite{Kim_2021} investigated how light is perturbed in the presence of GWs from a general relativistic perspective, by solving Maxwell’s equations 
in the geometrical-optics limit. 

In the aforementioned works, the interaction between EMWs and GWs is assumed to take place far away from the source of GWs; therefore, the strain amplitude of GWs can be treated as an extremely small constant value. 
In contrast with them, Cruise \cite{Cruise_1983} estimated the rotation angle of the plane of polarization of an EMW, which results from the passage of the wave through a non-flat spacetime with a GW present.
Here the EMW-GW interaction can occur close to the GW source and continues all along the EMW propagation path towards an observer. In this case, the strain amplitude can be initially large and decreases inversely 
as the distance from the source to a point of EMW-GW intersection; as a result, the cumulative effect from the EMW-GW interaction over the long propagation path can be outstanding in comparison with all the aforementioned 
cases, in which EMWs interact with plane GWs. However, despite a fresh perspective on the problem, Cruise \cite{Cruise_1983} overlooked some other important aspects: the spherical property of the GW, i.e., the dependence 
of the wave's amplitude and phase on $\left( r, \theta, \phi \right)$ in spherical polar coordinates, and its influence on EMW-GW interaction.\footnote{Instead, the GW is modeled in a cylindrical form.} Besides, he used several 
simplifying assumptions, with which to make his analysis easier, and this inevitably led to some loss of generality in his approach.

In this paper, we investigate the interaction between EMWs and GWs, and its effects on the properties of light, in a situation where light emitted from the neighborhood of a binary meets GWs from the binary 
all along the way to an observer. To this end, the paper proceeds largely in two steps as follows. In Sec. \ref{emw}, we will model the situation effectively using spherical GWs,\footnote{To be precise, `spherical GWs' 
refers to `transverse-traceless radially propagating GWs', as will be clarified in Sec. \ref{spacetime}, but we will use this term instead for convenience later.} and perform a geometrical-optics analysis in curved spacetime, 
with the curvature being due to the GWs propagating in flat spacetime. In Sec. \ref{appl}, for application of our analysis, we will focus in particular on the effect of gravitational Faraday rotation (or Skrotskii/Rytov effect), 
resulting from the interaction between EMWs and GWs from binaries, such as a supermassive BH binary in quasar and a double NS system.

\section{EMWs propagating through the GW background}
\label{emw}

\subsection{Spacetime with spherical GWs}
\label{spacetime}
Suppose a binary system of two masses $M_{1}$ and $M_{2}$ moving around each other in an orbit. Then transverse-traceless (TT) radially propagating GWs 
from this source are expressed as 
\begin{equation}
h_{ij}^{\mathrm{TT}}=\frac{2G}{c^{4}r}\left(P_{i}^{\, k}P_{j}^{\, l}-\frac{1}{2}P_{ij}P^{kl}\right)\ddot{Q}_{kl}\left(t_{\mathrm{R}}\right), \label{hTT}
\end{equation}
where the Latin indices $i$, $j$, $\ldots$ refer to the spatial coordinates, and $G\approx6.6743\times10^{-11}\,\mathrm{m^{3}\,kg^{-1}\,s^{-2}}$ is the gravitational constant, and
$c\approx2.998\times10^{8}\,\mathrm{m\,s^{-1}}$ is the speed of light, and $P^{ij}=\delta^{ij}-n^{i}n^{j}$ is the transverse projection tensor, and
\begin{equation}
Q^{ij}=\mu X^{i}X^{j}-\frac{1}{3}\delta^{ij}\left(\mu X^{k}X_{k}\right), \label{mq}
\end{equation}
denotes the mass quadrupole moment of the system, with $\mu\equiv M_{1}M_{2}/(M_{1}+M_{2})$ being the reduced mass and $\mathbf{X}\equiv\mathbf{r}_{1}-\mathbf{r}_{2}$ referring to 
the relative coordinates of the two bodies, and $t_{\mathrm{R}}=t-\mathbf{n}\cdot \mathbf{r}/c$ is the retarded time.

In a suitably chosen frame, the nonzero components of the quadrupole moment (\ref{mq}) are obtained for a circular orbit: 
\begin{eqnarray}
Q_{xx} &=&\mu R_{\mathrm{o}}^{2}\left[ \cos ^{2}\left( \Omega t_{\mathrm{R}}\right) -%
\frac{1}{3}\right] ,  \label{Qxx} \\
Q_{yy} &=&\mu R_{\mathrm{o}}^{2}\left[ \sin ^{2}\left( \Omega t_{\mathrm{R}}\right) -%
\frac{1}{3}\right] ,  \label{Qyy} \\
Q_{xy} &=&Q_{yx}=\mu R_{\mathrm{o}}^{2}\cos \left( \Omega t_{\mathrm{R}}\right) \sin
\left( \Omega t_{\mathrm{R}}\right) ,  \label{Qxy} \\
Q_{zz} &=&-\frac{\mu R_{\mathrm{o}}^{2}}{3},  \label{Qzz}
\end{eqnarray}
where $R_{\mathrm{o}}=\left\vert \mathbf{X} \right\vert$ is the radius of the orbit, and $\Omega =\sqrt{G(M_{1}+M_{2})/R_{\mathrm{o}}^{3}}$ is the orbital angular
velocity. Now, choosing $\mathbf{n}=\mathbf{e}_{r}=\left( \sin \theta \cos \phi ,\sin \theta \sin \phi ,\cos \theta \right)$ for $P^{ij}=\delta^{ij}-n^{i}n^{j}$ and 
$t_{\mathrm{R}}=t-\mathbf{n}\cdot \mathbf{r}/c$, one can work out all the components of the GWs from (\ref{hTT}) in the Cartesian coordinate frame 
$\left(\mathbf{e}_{x},\mathbf{e}_{y},\mathbf{e}_{z}\right)$:
\begin{align}
h_{xx}^{\mathrm{TT}} & = \mathcal{H}\left(r\right)\left[-\cos^{2}\theta\cos\left(-\omega_{\mathrm{g}}t+kr\right)
+\frac{1}{2}\sin^{2}\theta\left(1-\sin^{2}\theta\cos^{2}\phi\right)\cos\left(-\omega_{\mathrm{g}}t+kr+2\phi\right)\right], \label{hxx} \\
h_{xy}^{\mathrm{TT}} & = \mathcal{H}\left(r\right)\left[\cos^{2}\theta\sin\left(-\omega_{\mathrm{g}}t+kr\right)
-\frac{1}{4}\sin^{4}\theta\sin\left(2\phi\right)\cos\left(-\omega_{\mathrm{g}}t+kr+2\phi\right)\right], \label{hxy} \\
h_{xz}^{\mathrm{TT}} & = \mathcal{H}\left(r\right)\left[\sin\theta\cos\theta\cos\left(-\omega_{\mathrm{g}}t+kr+\phi\right)
-\frac{1}{2}\sin^{3}\theta\cos\theta\cos\phi\cos\left(-\omega_{\mathrm{g}}t+kr+2\phi\right)\right], \label{hxz} \\
h_{yy}^{\mathrm{TT}} & = \mathcal{H}\left(r\right)\left[\cos^{2}\theta\cos\left(-\omega_{\mathrm{g}}t+kr\right)
+\frac{1}{2}\sin^{2}\theta\left(1-\sin^{2}\theta\sin^{2}\phi\right)\cos\left(-\omega_{\mathrm{g}}t+kr+2\phi\right)\right], \label{hyy} \\
h_{yz}^{\mathrm{TT}} & = \mathcal{H}\left(r\right)\left[-\sin\theta\cos\theta\sin\left(-\omega_{\mathrm{g}}t+kr+\phi\right)
-\frac{1}{2}\sin^{3}\theta\cos\theta\sin\phi\cos\left(-\omega_{\mathrm{g}}t+kr+2\phi\right)\right], \label{hyz} \\
h_{zz}^{\mathrm{TT}} & = \mathcal{H}\left(r\right)\left[-\frac{1}{2}\sin^{2}\theta\left(1+\cos^{2}\theta\right)\cos\left(-\omega_{\mathrm{g}}t+kr+2\phi\right)\right], \label{hzz}
\end{align}
where $\mathcal{H}\left(r\right) \equiv G\mu R_{\mathrm{o}}^{2}\omega_{\mathrm{g}}^{2}/\left(c^{4}r\right)$ represents the \textit{radially varying} strain amplitude
and $\omega_{\mathrm{g}} \equiv 2\Omega$ denotes the GW frequency.

Now, the spacetime geometry, in which Maxwell's equations are solved, reads 
\begin{align}
ds^{2} = &-c^{2}dt^{2}+\left(1+h_{xx}^{\mathrm{TT}}\right)dx^{2}+2h_{xy}^{\mathrm{TT}}dxdy+2h_{xz}^{\mathrm{TT}}dxdz \notag \\
&+\left(1+h_{yy}^{\mathrm{TT}}\right)dy^{2}+2h_{yz}^{\mathrm{TT}}dydz+\left(1+h_{zz}^{\mathrm{TT}}\right)dz^{2}, \label{ds}
\end{align}
where the components of $h_{ij}^{\mathrm{TT}}$ refer to (\ref{hxx})-(\ref{hzz}). Then it is implied that our solutions to the Maxwell's equations defined in this spacetime geometry naturally 
contain the interaction between EMWs and GWs.

One should note that transforming the GWs as given by (\ref{hxx})-(\ref{hzz}) above from the Cartesian coordinate frame $\left(\mathbf{e}_{x},\mathbf{e}_{y},\mathbf{e}_{z}\right)$ 
to the spherical coordinate frame $(\mathbf{e}_{r},\mathbf{e}_{\hat{\theta}},\mathbf{e}_{\hat{\phi}})$ yields the compact expressions of `spherical' GWs from the binary source 
\cite{Maggiore_2007,thorne2017modern}:  
\begin{align}
h_{+}\left(t,r,\theta,\phi\right) &\equiv h_{\hat{\theta}\hat{\theta}}^{\mathrm{TT}}=-h_{\hat{\phi}\hat{\phi}}^{\mathrm{TT}} \notag \\
&= -\mathcal{H}\left(r\right)\left(\frac{1+\cos^{2}\theta}{2}\right)\cos\left(-\omega_{\mathrm{g}}t+kr+2\phi\right), \label{hp} \\
h_{\times}\left(t,r,\theta,\phi\right) &\equiv h_{\hat{\theta}\hat{\phi}}^{\mathrm{TT}}=h_{\hat{\phi}\hat{\theta}}^{\mathrm{TT}} \notag \\
&= \mathcal{H}\left(r\right)\cos\theta\sin\left(-\omega_{\mathrm{g}}t+kr+2\phi\right), \label{hc}
\end{align}
where the subscript indices $\hat{\theta}$ and $\hat{\phi}$ denote the basis vectors $\mathbf{e}_{\hat{\theta}}$ and $\mathbf{e}_{\hat{\phi}}$, respectively.

\subsection{Geometrical-optics analysis for EMWs in the GW background}
\label{gmo}
Light (EMWs) propagating in a spacetime curved due to GWs, as given in Sec. \ref{spacetime}, can be described by Maxwell's equations:
\begin{equation}
\square A^{\mu}-R^{\mu}{}_{\nu}A^{\nu}=0, \label{me}
\end{equation}
where the Greek indices $\mu$, $\nu$, $\ldots$ refer to the temporal and spatial coordinates, and $\square A^{\mu} \equiv g^{\nu\rho}\nabla_{\nu}\nabla_{\rho}A^{\mu}$ means the d’Alembertian 
on a vector potential and, $R_{\mu\nu}$ denotes the Ricci tensor. Here both the operator and the tensor are defined in the spacetime geometry as given by Eq. (\ref{ds}), and therefore expressed 
in terms of $h_{ij}^{\mathrm{TT}}$. However, in Lorenz gauge ($\partial^{\mu}\bar{h}_{\mu\nu}=0$) and TT gauge ($h_{0\mu}=0$ and $h_{\,\,\,\mu}^{\mu}=0$), it turns out
\begin{equation}
R_{\mu\nu}=\mathcal{O}\left(h^{2}\right). \label{ricci}
\end{equation}
Therefore, the Maxwell's equations (\ref{me}) reduce to
\begin{equation}
\square A^{\mu}=\mathcal{O}\left(h^{2}\right). \label{redme}
\end{equation}

We consider a geometrical-optics ansatz for Eq. (\ref{redme}) \cite{poisson2014gravity,dolan2018geometrical}:
\begin{equation}
A^{\mu}\left(x\right)=\mathcal{A}\left(x\right)\varepsilon^{\mu}\left(x\right)\exp\left(\mathrm{i}\varpi\Psi\left(x\right)\right), \label{goans}
\end{equation}
where $\mathcal{A}\left(x\right)$ is the amplitude, $\varepsilon^{\mu}\left(x\right)$ denotes the polarization vector, and $\exp\left(\mathrm{i}\varpi\Psi\left(x\right)\right)$ represents 
the wave field, with $\varpi$ being a dimensionless parameter.
Plugging this into the Maxwell's equations (\ref{redme}), we obtain 
\begin{eqnarray}
(i) &  & -{\,{\varpi^{2}\left(K^{\nu}K_{\nu}\mathcal{A}\varepsilon^{\mu}\right)}}\notag \protect\\
{(ii)} &  & {+\,\mathrm{i}{\varpi^{1}}\left[\varepsilon^{\mu}{\nabla_{\nu}\left(\mathcal{A}^{2}K^{\nu}\right)}+2\mathcal{A}{K^{\nu}\nabla_{\nu}\varepsilon^{\mu}}\right]}\notag \protect\\
(iii) &  & +\,\mathcal{O}\left(\varpi^{0}\right)=0, \label{goans1}
\end{eqnarray}
where $\varpi$ serves as an order-counting parameter and $K^{\mu}\equiv\nabla^{\mu}\Psi$ denotes the wave vector of EMWs.

From the null condition $K^{\mu}K_{\mu}=0$ (which makes the term (\textit{i}) in Eq. (\ref{goans1}) vanish) and with the term (\textit{ii}) in Eq. (\ref{goans1}) vanishing, we find 
\begin{eqnarray}
{K^{\nu}\nabla_{\nu}K^{\mu}=0} &  & (\mathrm{geodesic\; equation}), \label{goans2} \\ 
{\nabla_{\nu}\left(\mathcal{A}^{2}K^{\nu}\right)=0} &  & (\mathrm{conservation\; of\; flux}), \label{goans3} \\
{K^{\nu}\nabla_{\nu}\varepsilon^{\mu}=0} &  & (\mathrm{Skrotskii/Rytov\; effect}). \label{goans4}
\end{eqnarray}
Solving these equations, we obtain 
\begin{align}
{\delta K_{[1]}^{\mu}} & \simeq {-K_{\left[0\right]}^{\nu}\int(K_{\left[0\right]}^{\rho}k_{\rho}){\tilde{h}^{\mathrm{s}}{}_{\,\,\,\nu}^{\mu}}\mathrm{d}\lambda
+\frac{K_{\left[0\right]}^{\nu}K_{\left[0\right]}^{\rho}}{2}\int k^{\mu}{\tilde{h}_{\nu\rho}^{\mathrm{s}}}\mathrm{d}\lambda}, \label{goans5} \\
{\delta\Psi_{[1]}} & \simeq {-\frac{K_{\left[0\right]}^{\mu}K_{\left[0\right]}^{\nu}}{2}\int\left[\int(K_{\left[0\right]}^{\rho}k_{\rho}){\tilde{h}_{\mu\nu}^{\rm{s}}}\mathrm{d}\lambda\right]
\mathrm{d}\lambda}, \label{goans6} \\
{\delta\mathcal{A}_{[1]}} & \simeq {0}, \label{goans7} \\
{\delta\varepsilon_{[1]}^{\mu}} & \simeq {-\frac{\varepsilon_{\left[0\right]}^{\nu}}{2}\int(K_{\left[0\right]}^{\rho}k_{\rho}){\tilde{h}^{\rm{s}}{}_{\,\,\,\nu}^{\mu}}\mathrm{d}\lambda
-\frac{K_{\left[0\right]}^{\nu}}{2}\int(\varepsilon_{\left[0\right]}^{\rho}k_{\rho}){\tilde{h}^{\mathrm{s}}{}_{\,\,\,\nu}^{\mu}}\mathrm{d}\lambda
+\frac{K_{\left[0\right]}^{\nu}\varepsilon_{\left[0\right]}^{\rho}}{2}\int k^{\mu}{\tilde{h}_{\nu\rho}^{\mathrm{s}}}\mathrm{d}\lambda}, \label{goans8} 
\end{align}
where we have assumed a perturbative solution for each quantity, i.e., $K^{\mu}=K_{[0]}^{\mu}+\delta K_{[1]}^{\mu}+\mathcal{O}\left(h^{2}\right)$, 
$\Psi=\Psi_{[0]}+\delta\Psi_{[1]}+\mathcal{O}\left(h^{2}\right)$ (\emph{c.f.} $K_{[0]}^{\mu}=\nabla^{\mu}\Psi_{[{0}]}$), 
$\mathcal{A}=\mathcal{A}_{[{0}]}+\delta\mathcal{A}_{[{1}]}+\mathcal{O}\left(h^{2}\right)$
and $\varepsilon^{\mu}=\varepsilon_{[{0}]}^{\mu}+\delta\varepsilon_{[{1}]}^{\mu}+\mathcal{O}\left(h^{2}\right)$,
with the subscripts $[0]$ and $[1]$ meaning `unperturbed' and `first-order in $h$', respectively,
and $\lambda$ is an affine parameter defined via $\mathrm{d}x^{\mu}/\mathrm{d}\lambda=K_{\left[{0}\right]}^{\mu}$.
Here we have renamed and redefined $h_{\mu\nu}^{\mathrm{TT}}$ from Eqs. (\ref{hxx})-(\ref{hzz}):
\begin{align}
h_{\mu\nu}^{\mathrm{s}} &=\Re\left(\mathfrak{H}_{\mu\nu}\left(r,\theta,\phi\right)\exp\left(\mathrm{i}k_{\alpha}x^{\alpha}\right)\right), \label{hmn} \\
\tilde{h}_{\mu\nu}^{\mathrm{s}} &=\Re\left(\mathrm{i}\mathfrak{H}_{\mu\nu}\left(r,\theta,\phi\right)\exp\left(\mathrm{i}k_{\alpha}x^{\alpha}\right)\right), \label{h_mn}
\end{align}
where the label `$\mathrm{TT}$' (meaning \textit{transverse-traceless}) has been replaced by `$\mathrm{s}$' (meaning \textit{spherical}), and $\mathfrak{H}_{\mu\nu}\left(r,\theta,\phi\right)$ 
represents the complex strain tensor, which can be read off from (\ref{hxx})-(\ref{hzz}). Note that `$\simeq$' signs in (\ref{goans5})-(\ref{goans8}) above are due to 
$\partial^{\mu}h_{\nu\rho}^{\mathrm{s}}\simeq k^{\mu}\tilde{h}_{\nu\rho}^{\mathrm{s}}$.\footnote{One can deduce from (\ref{hxx})-(\ref{hzz}), (\ref{hmn}) and (\ref{h_mn}), 
$\partial ^{\mu }h_{\nu \rho }^{\mathrm{s}}=k^{\mu}\tilde{h}_{\nu \rho}^{\mathrm{s}}+\mathcal{O}\left( r^{-1}\mathcal{H}\right)$, where the first term on the right-hand side comes from 
$\Re \left( \mathfrak{H}_{\nu \rho}\partial^{\mu}\exp \left[ \mathrm{i}\left(k_{\sigma }x^{\sigma}\right) \right] \right) $, while the second term is due to 
$\Re \left( \left(\partial^{\mu}\mathfrak{H}_{\nu \rho}\right) \exp \left[ \mathrm{i}\left(k_{\sigma }x^{\sigma }\right) \right] \right) $. Here we see
$\partial^{\mu}\mathfrak{H}_{\nu \rho}\sim \partial^{\mu}\left[ \left(1/r\right) \left( A_{\nu \rho}^{ij}x^{i}x^{j}/r^{2} + B_{\nu \rho}^{ijkl}x^{i}x^{j}x^{k}x^{l}/r^{4} \right) \right]$, 
with the coefficients $A_{\nu \rho}^{ij}$ and $B_{\nu \rho}^{ijkl}$ being of the same dimension as $G\mu R_{\mathrm{o}}^{2}\omega _{\mathrm{g}}^{2}/c^{4}=r\mathcal{H}$ (i.e., length), 
and therefore $\Re \left(\left(\partial ^{\mu}\mathfrak{H}_{\nu \rho}\right) \exp\left[\mathrm{i}\left(k_{\sigma}x^{\sigma}\right) \right] \right)
\sim \mathcal{O}\left( r^{-1}\mathcal{H}\right) \ll \mathcal{O}\left(\mathcal{H}^{2}\right)$, which can be disregarded in the linearized gravity limit. 
However, in order for this approximation to be valid, we require $k=\omega _{\mathrm{g}}/c\gg r^{-1}$, which restricts $r_{\min }\gg c/\omega _{\mathrm{g}}$; 
e.g., for GWs with $\omega _{\mathrm{g}}\sim 10^{-7}\,\mathrm{Hz}$ (from a supermassive black hole binary as presented in Example 1 in Sec. \ref{fr}), 
$r_{\min }\gg 10^{15}\,\mathrm{m}\sim 0.1\,\mathrm{pc}$.\label{fn3}}

Finally, our solutions to Eq. (\ref{redme}) are given by means of (\ref{goans}), in a perturbative form:
\begin{equation}
A^{\mu}\left(x\right)=\mathcal{A}_{[0]}\left[\varepsilon_{[0]}^{\mu}+{\delta\varepsilon_{[1]}^{\mu}\left(x\right)}\right]
\exp\left( \mathrm{i}\left(K_{\alpha[{0}]}x^{\alpha}+\delta\Psi_{[1]}\left(x\right) \right) \right), \label{ansatz}
\end{equation}
where $\delta\Psi_{[1]}$ and $\delta\varepsilon_{[1]}^{\mu}$ refer to Eqs. (\ref{goans6}) and (\ref{goans8}), respectively. In reference to (\ref{ansatz}) together with (\ref{goans5})-(\ref{goans8}), 
one should note the following: based on the geometrical-optics analysis, we have first-order modulations of the phase, the wave vector and the polarization vector of EMWs due to GWs, 
with the exception of the amplitude, and our solutions to the Maxwell's equations are expressed in terms of such modulations.

\section{Applications for EMW-GW interaction: (1) Supermassive BH binary in quasar, (2) Double neutron star system}
\label{appl}

\subsection{Binaries and GWs}
\label{bgw}
As a good example for the EMW-GW interaction, one can consider quasars, the brilliant cores of active galaxies, as they may commonly host two central supermassive BHs, which fall into an orbit 
about one another as a result of the merger between two galaxies; e.g., Markarian 231 (Mrk 231), the nearest galaxy to Earth that hosts a quasar, has been found to be powered by two such BHs, 
using NASA’s Hubble Space Telescope. The binary BHs generate a tremendous amount of energy that makes the core of the host galaxy outshine the glow of its population of billions of stars 
\cite{Yan_2015,Hubble2015}. At the same time, light from a quasar will interact with GWs emitted by the binary. In this situation, however, the GW strain amplitude $h$ is not treated as a constant value; 
rather, it varies with the distance $r$ from the source to a field point where the two wavefronts of the light and the GWs meet each other. Therefore, the interaction between the light and the GWs 
weakens as $h$ decreases over the distance, but accumulates over the entire path of light propagation.

Neutron star (NS) binaries are another example we can take from which to investigate the EMW-GW interaction. For instance, PSR J1913+16 is known as the first double neutron star (DNS) system, 
discovered by Hulse and Taylor in 1974 \cite{Hulse_1975}. This binary system consists of a pulsar and a massive companion orbiting around each other, gravitational interactions of which can be 
indirectly verified via its orbital contraction, caused by the energy loss of the system emitting GWs. There are 19 DNS systems that have been discovered until now (as of 2020), and amongst them, 
PSR J0737–3039 is the only one known to possess both the primary and the companion NSs detected as pulsars \cite{Burgay_2003,Lyne_2004,Liu_2021}. Likewise, as $h$ decreases over the distance, 
the interaction between the pulsar emissions and the GWs weakens, but accumulates over the entire path of light propagation.

\subsection{Faraday rotation (Skrotskii/Rytov effect) by GWs}
\label{fr}

\begin{figure}[!ht]
\centering
\protect\includegraphics[scale=0.5]{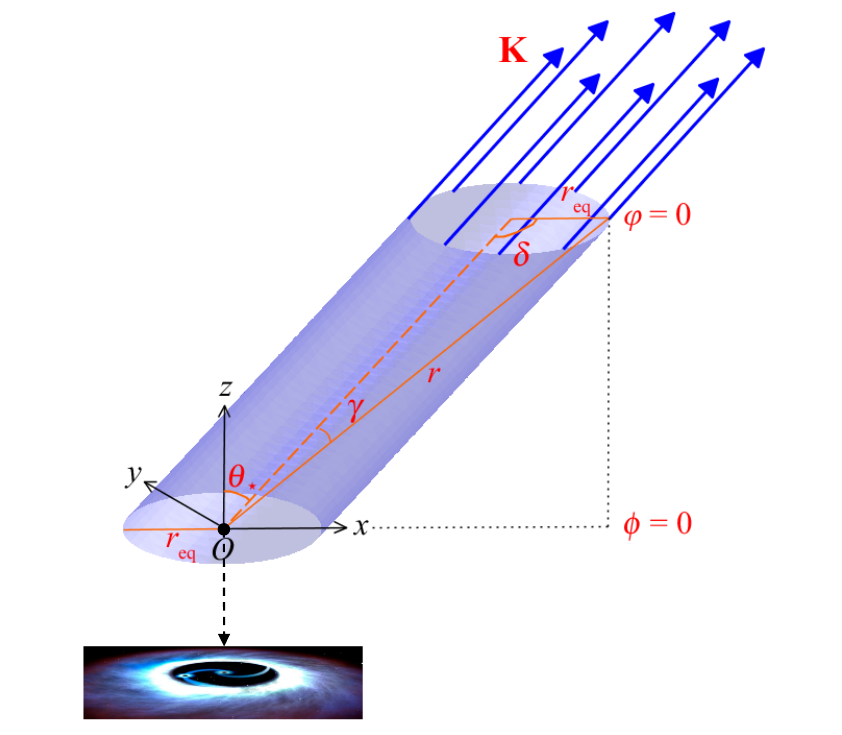}
\caption{Oblique cylinder: $\left(x-z\tan\theta_{\star}\right)^{2}+y^{2}=r_{\mathrm{eq}}^{2}$. A black hole binary as a source of GWs is located at the origin $O$. 
Consider a circular ring of radius $r_{\mathrm{eq}}$ encircling the binary on the equatorial plane, from which light beams propagate along the direction, 
tilted by an angle $\theta_{\star}$ from the normal to the plane. The interaction between the quasar's light and the GWs can be described as occurring 
on the surface of an oblique cylinder with the radius $r_{\mathrm{eq}}$ and the tilt angle $\theta_{\star}$, where $r_{\mathrm{eq}}$ equals the initial distance 
from the GW source to a photon on the equatorial plane. The size of $r_{\mathrm{eq}}$ is considered to range from the core of the active galactic nucleus
($\sim1\:\mathrm{pc}$) to the molecular disk ($\sim100\:\mathrm{pc}$). The weak-field perturbations via the GW strain amplitude are valid as this region 
is far enough from the binary to ignore the gravitational effect on it; $r_{\mathrm{eq}}\gg R_{\mathrm{o}}$ (binary radius). (Credit: \cite{Hubble2015} for the illustration of a binary BH at the bottom.)}
\label{fig_cyl}
\end{figure}

Among the effects due to GWs, we focus on the Faraday rotation of the polarization vector of EMWs. Recalling from Sec. \ref{gmo}, the first-order modulation of the polarization vector due to GWs 
is given by (\ref{goans8}), which is a solution to (\ref{goans4}); that is, 
\begin{align}
K^{\nu}\nabla_{\nu}\varepsilon^{\mu} &= 0\;\;\;(\mathrm{Skrotskii/Rytov\; effect}) \notag \\
\Rightarrow \;\; {\delta\varepsilon_{[{1}]}^{\mu}} &\simeq {-\frac{\varepsilon_{\left[{0}\right]}^{\nu}}{2}\int \left(K_{\left[{0}\right]}^{\rho}k_{\rho}\right){\tilde{h}^{\rm{s}}{}_{\,\,\,\nu}^{\mu}}
\mathrm{d}\lambda -\frac{K_{\left[{0}\right]}^{\nu}}{2}\int \left(\varepsilon_{\left[{0}\right]}^{\rho}k_{\rho}\right){\tilde{h}^{\rm{s}}{}_{\,\,\,\nu}^{\mu}}\mathrm{d}\lambda
+\frac{K_{\left[{0}\right]}^{\nu}\varepsilon_{\left[{0}\right]}^{\rho}}{2}\int k^{\mu}{\tilde{h}_{\nu\rho}^{\mathrm{s}}}\mathrm{d}\lambda}. \label{dep}
\end{align}
Then using this, the Faraday rotation angle $\varphi _{\mathrm{F}}$ can be computed as
\begin{equation}
\varphi _{\mathrm{F}} = \sin ^{-1}\left(\left( \boldsymbol{\varepsilon}_{\left[0\right] }\times \delta\boldsymbol{\varepsilon}_{[{1}]}\right)\cdot \mathbf{\hat{K}}_{\left[ 0\right]} \right)
\approx \left( \boldsymbol{\varepsilon}_{\left[0\right] }\times \delta\boldsymbol{\varepsilon}_{[{1}]}\right)\cdot \mathbf{\hat{K}}_{\left[ 0\right]} + \mathcal{O}\left(h^2 \right),  \label{fra}
\end{equation}%
where $\mathbf{\hat{K}}_{\left[ 0\right]}\equiv \mathbf{K}_{\left[ 0\right]}/\lvert \mathbf{K}_{\left[ 0\right]} \rvert = c\mathbf{K}_{\left[ 0\right]}/\omega _{\mathrm{e}}$ denotes 
the unit spatial propagation vector of our quasar's light, with $\omega _{\mathrm{e}}$ being the light frequency. The direction of $\mathbf{\hat{K}}_{\left[ 0\right]}$ is represented by the orange dashed line 
in Fig. \ref{fig_cyl}. 

Now, for simplification, one may set $\phi_{\star}=0$ for $\mathbf{\hat{K}}_{\left[ 0\right]}$ as represented by the two angles $\left( \theta _{\star },\phi _{\star }\right) $, without loss of generality 
(see Fig. \ref{fig_cyl}):
\begin{equation}
\mathbf{\hat{K}}_{\left[ 0\right]}=\left( \sin \theta_{\star} \cos \phi_{\star}, \sin \theta_{\star} \sin \phi_{\star}, \cos \theta_{\star} \right)
=\left( \sin \theta_{\star},0,\cos \theta_{\star} \right).  \label{K0}
\end{equation}%
In the Coulomb gauge, however, one can consider a pair of polarization vectors associated with this:\footnote{Note that the three vectors, $\mathbf{\hat{K}}_{\left[ 0\right]}$, 
$\boldsymbol{\varepsilon}_{\mathrm{I}\left[ 0\right]}$ and $\boldsymbol{\varepsilon}_{\mathrm{II}\left[ 0\right]}$ form an orthonormal basis} 
\begin{align}
\boldsymbol{\varepsilon}_{\mathrm{I}\left[ 0\right]} &= \left( \cos \theta_{\star}, 0, -\sin \theta_{\star} \right),  \label{eps01} \\
\boldsymbol{\varepsilon}_{\mathrm{II}\left[ 0\right]} &= \left( 0, 1, 0 \right).  \label{eps02}
\end{align}

By means of Eqs. (\ref{dep})-(\ref{eps02}) one can determine the Faraday rotation angles, for the polarization vectors $\boldsymbol{\varepsilon}_{\mathrm{I}\left[ 0\right]}$ and 
$\boldsymbol{\varepsilon}_{\mathrm{II}\left[ 0\right]}$, respectively as
\begin{align}
\varphi _{\mathrm{F,I}} &\approx \frac{\cos\theta_{\star}}{2}\int^{\lambda_{\mathrm{f}}}_{\lambda_{\mathrm{i}}}\left(K_{\left[{0}\right]}^{\mu}k_{\mu}\right)
\tilde{h}_{xy}^{\mathrm{s}} \mathrm{d}\lambda -\frac{\sin\theta_{\star}}{2}\int^{\lambda_{\mathrm{f}}}_{\lambda_{\mathrm{i}}}\left(K_{\left[{0}\right]}^{\mu}k_{\mu}\right)
\tilde{h}_{yz}^{\mathrm{s}} \mathrm{d}\lambda \notag \\
&\hspace{12pt}+\frac{\omega_{\mathrm{e}}}{c}\left[\frac{\sin\left(2\theta_{\star}\right)}{4}\int^{\lambda_{\mathrm{f}}}_{\lambda_{\mathrm{i}}}\left(k_{x}\tilde{h}_{xy}^{\mathrm{s}} -k_{z}\tilde{h}_{yz}^{\mathrm{s}} -k_{y}\tilde{h}_{xx}^{\mathrm{s}}
+k_{y}\tilde{h}_{zz}^{\mathrm{s}}\right)\mathrm{d}\lambda -\frac{\cos\left(2\theta_{\star}\right)}{2}\int^{\lambda_{\mathrm{f}}}_{\lambda_{\mathrm{i}}}k_{y}\tilde{h}_{xz}^{\mathrm{s}}
\mathrm{d}\lambda\right. \notag \\
&\hspace{12pt}\left.+\frac{\cos^{2}\theta_{\star}}{2}\int^{\lambda_{\mathrm{f}}}_{\lambda_{\mathrm{i}}}k_{x}\tilde{h}_{yz}^{\mathrm{s}}\mathrm{d}\lambda -\frac{\sin^{2}\theta_{\star}}{2}\int^{\lambda_{\mathrm{f}}}_{\lambda_{\mathrm{i}}}k_{z}\tilde{h}_{xy}^{\mathrm{s}}\mathrm{d}\lambda \right] +\mathcal{O}\left(h^2 \right), \label{fra1} \\
\varphi _{\mathrm{F,II}} &\approx -\frac{\cos\theta_{\star}}{2}\int^{\lambda_{\mathrm{f}}}_{\lambda_{\mathrm{i}}}\left(K_{\left[{0}\right]}^{\mu}k_{\mu}\right)
\tilde{h}_{xy}^{\mathrm{s}} \mathrm{d}\lambda +\frac{\sin\theta_{\star}}{2}\int^{\lambda_{\mathrm{f}}}_{\lambda_{\mathrm{i}}}\left(K_{\left[{0}\right]}^{\mu}k_{\mu}\right)
\tilde{h}_{yz}^{\mathrm{s}} \mathrm{d}\lambda \notag \\
&\hspace{12pt}+\frac{\omega_{\mathrm{e}}}{c}\left[\frac{\sin\left(2\theta_{\star}\right)}{4}\int^{\lambda_{\mathrm{f}}}_{\lambda_{\mathrm{i}}}\left(k_{x}\tilde{h}_{xy}^{\mathrm{s}} -k_{z}\tilde{h}_{yz}^{\mathrm{s}} -k_{y}\tilde{h}_{xx}^{\mathrm{s}}
+k_{y}\tilde{h}_{zz}^{\mathrm{s}}\right)\mathrm{d}\lambda -\frac{\cos\left(2\theta_{\star}\right)}{2}\int^{\lambda_{\mathrm{f}}}_{\lambda_{\mathrm{i}}}k_{y}\tilde{h}_{xz}^{\mathrm{s}}
\mathrm{d}\lambda\right. \notag \\
&\hspace{12pt}\left.+\frac{\cos^{2}\theta_{\star}}{2}\int^{\lambda_{\mathrm{f}}}_{\lambda_{\mathrm{i}}}k_{x}\tilde{h}_{yz}^{\mathrm{s}}\mathrm{d}\lambda -\frac{\sin^{2}\theta_{\star}}{2}\int^{\lambda_{\mathrm{f}}}_{\lambda_{\mathrm{i}}}k_{z}\tilde{h}_{xy}^{\mathrm{s}}\mathrm{d}\lambda \right] +\mathcal{O}\left(h^2 \right), \label{fra2} 
\end{align}
where 
\begin{equation}
\mathbf{k}=\left(k_{x},k_{y},k_{z}\right)=\frac{\omega_{\mathrm{g}}}{c}\left( \sin \theta \cos \phi ,\sin \theta \sin \phi ,\cos \theta \right), \label{k}
\end{equation}
and 
\begin{equation}
K_{\left[{0}\right]}^{\mu}k_{\mu}=-\frac{\omega _{\mathrm{e}}\omega _{\mathrm{g}}}{c^2}+\mathbf{K}_{\left[ 0\right]}\cdot\mathbf{k}
=-\frac{\omega _{\mathrm{e}}\omega _{\mathrm{g}}}{c^2}\left[1-\left(\cos \theta \cos \theta _{\star }+\sin \theta \sin \theta_{\star }\cos \phi\right)\right], \label{Kk4}
\end{equation}
and in view of (\ref{hxx})-(\ref{hzz}), (\ref{hmn}) and (\ref{h_mn}) the components of GWs can be written out:
\begin{align}
{{{\tilde{h}_{xx}^{\mathrm{s}}}}} & {=}  {\mathcal{H}\left(r\right)\left[\cos^{2}\theta\sin\left(-\omega_{\mathrm{g}}t+kr\right)
-\frac{1}{2}\sin^{2}\theta\left(1-\sin^{2}\theta\cos^{2}\phi\right)\sin\left(-\omega_{\mathrm{g}}t+kr+2\phi\right)\right]}, \label{h_xx} \\
{{{\tilde{h}_{xy}^{\mathrm{s}}}}} & {=}  {\mathcal{H}\left(r\right)\left[-\cos^{2}\theta\cos\left(-\omega_{\mathrm{g}}t+kr\right)
+\frac{1}{4}\sin^{4}\theta\sin\left(2\phi\right)\sin\left(-\omega_{\mathrm{g}}t+kr+2\phi\right)\right]}, \label{h_xy} \\
{{{\tilde{h}_{xz}^{\mathrm{s}}}}} & {=}  {\mathcal{H}\left(r\right)\left[-\sin\theta\cos\theta\sin\left(-\omega_{\mathrm{g}}t+kr+\phi\right)
+\frac{1}{2}\sin^{3}\theta\cos\theta\cos\phi\sin\left(-\omega_{\mathrm{g}}t+kr+2\phi\right)\right]}, \label{h_xz} \\
{{{\tilde{h}_{yz}^{\mathrm{s}}}}} & {=}  {\mathcal{H}\left(r\right)\left[\sin\theta\cos\theta\cos\left(-\omega_{\mathrm{g}}t+kr+\phi\right)
+\frac{1}{2}\sin^{3}\theta\cos\theta\sin\phi\sin\left(-\omega_{\mathrm{g}}t+kr+2\phi\right)\right]}, \label{h_yz} \\
{{{\tilde{h}_{zz}^{\mathrm{s}}}}} & {=}  {\mathcal{H}\left(r\right)\left[\frac{1}{2}\sin^{2}\theta\left(1+\cos^{2}\theta\right)\sin\left(-\omega_{\mathrm{g}}t+kr+2\phi\right)\right]}, \label{h_zz} 
\end{align}
with $\mathcal{H}\left(r\right)=G\mu R_{\mathrm{o}}^{2}\omega_{\mathrm{g}}^{2}/\left(c^{4}r\right)$.

To evaluate the Faraday rotation angles as given by Eqs. (\ref{fra1}) and (\ref{fra2}), it is convenient to have the following quantities, all expressed in terms of a single parameter $\xi\equiv\tan\left(\gamma/2\right)$, 
where $\gamma$ denotes the angle between the propagation directions of light and GWs, defined via $\cos\gamma\equiv\cos \theta \cos \theta _{\star }+\sin \theta \sin \theta_{\star }\cos \phi$ 
(see Fig. \ref{fig_cyl} and \ref{appA} for details):
\begin{align}
r &=\frac{r_{\mathrm{eq}}\sqrt{1-\sin^{2}\theta_{\star}\cos^{2}\varphi}\left( 1+\xi^{2} \right)}{2\xi},\label{r} \\
\sin\theta\cos\phi &= {\frac{2\cos^{2}\theta_{\star}\cos\varphi\xi}{\sqrt{1-\sin^{2}\theta_{\star}\cos^{2}\varphi}\left(1+\xi^{2}\right)}
+\frac{\sin\theta_{\star}\left( 1-\xi^{2} \right)}{1+\xi^{2}}},\label{sthcph} \\
\sin\theta\sin\phi &= {\frac{2\sin\varphi\xi}{\sqrt{1-\sin^{2}\theta_{\star}\cos^{2}\varphi}\left(1+\xi^{2}\right)}},\label{sthsph} \\
\cos\theta &= {\frac{\cos\theta_{\star}\left(1-\xi^{2}\right)}{1+\xi^{2}}
-\frac{2\cos\theta_{\star}\sin\theta_{\star}\cos\varphi\xi}{\sqrt{1-\sin^{2}\theta_{\star}\cos^{2}\varphi}\left(1+\xi^{2}\right)}},\label{cth} \\
K_{\left[{0}\right]}^{\mu}k_{\mu} &= -\frac{2\omega_{\mathrm{e}}\omega_{\mathrm{g}}\xi^{2}}{c^{2}\left(1+\xi^{2}\right)},\label{Kdk} \\
\left. \left(-\omega_{\mathrm{g}}t+kr\right)\right\vert_{\mathrm{photon\, path}} 
&= -\omega_{\mathrm{g}}t_{0}+\frac{\omega_{\mathrm{g}}r_{\mathrm{eq}} \sqrt{1-\sin^{2}\theta_{\star}\cos^{2}\varphi}\xi}{c}, \label{phase} \\
\mathrm{d}\lambda &= \frac{\left.\mathrm{d}\left( k_{\mu }x^{\mu }\right)\right\vert_{\mathrm{photon\,path}}}{K_{\left[ {0}\right]}^{\nu}k_{\nu}} 
=-\frac{c r_{\mathrm{eq}}\sqrt{1-\sin^{2}\theta_{\star}\cos^{2}\varphi}\left(1+\xi^{2}\right)}{2\omega_{\mathrm{e}}\xi ^{2}}\mathrm{d}\xi, \label{dlmbd}
\end{align}
where the expressions in Eqs. (\ref{r})-(\ref{cth}) have been derived from the oblique cylinder as depicted in Fig. \ref{fig_cyl}, using the relations $\cos\gamma=\left(1-\xi^{2}\right)/\left(1+\xi^{2}\right)$ and 
$\sin\gamma=2\xi/\left(1+\xi^{2}\right)$, and in Eqs. (\ref{phase}) and (\ref{dlmbd}) $t=t_{0}+r\cos\gamma/c$ as defined along the photon path, and due to the substitution 
$\mathrm{d}\lambda\rightarrow\mathrm{d}\xi$ in Eq. (\ref{dlmbd}), the limits of integration in Eqs. (\ref{fra1}) and (\ref{fra2}) change: $\lambda_{\mathrm{i}}\rightarrow\xi_{\mathrm{i}}
=\sqrt{\left(1-\sin\theta_{\star}\cos\varphi\right)/\left(1+\sin\theta_{\star}\cos\varphi\right)}$ and $\lambda_{\mathrm{f}}\rightarrow\xi_{\mathrm{f}}=0$.

In consideration of Eqs. (\ref{fra1}) and (\ref{fra2}) with Fig. \ref{fig_cyl}, we investigate how the effect of the Faraday rotation depends upon the direction of a quasar's light beam, described by $\theta_{\star}$, with the condition for optimization, {$\varphi=0$}: 
\begin{equation}
\vartheta_{\mathrm{F,I/II}}\left(\theta_{\star};\varphi=0\right) \approx \frac{G\mu R_{\mathrm{o}}^{2}\omega_{\mathrm{g}}^{3}}{c^{5}}\times \mathcal{Q}_{\mathrm{I/II}}\left(\theta_{\star};\varphi=0\right), \label{fra3} 
\end{equation}
where the subscripts $\mathrm{I}$ and $\mathrm{II}$ refer to the polarization vectors $\boldsymbol{\varepsilon}_{\mathrm{I}\left[ 0\right]}$ and $\boldsymbol{\varepsilon}_{\mathrm{II}\left[ 0\right]}$, as given by (\ref{eps01}) and (\ref{eps02}), respectively, and
\begin{align}
\mathcal{Q}_{\mathrm{I}}\left(\theta_{\star};\varphi=0\right) &\equiv \Re\left\{\int_{\xi_{\mathrm{i}}}^{\xi_{\mathrm{f}}}d\xi\,
\frac{e^{\mathrm{i}\left(a\xi+b\right)}}{\left(1+\xi^{2}\right)^{3}}\left( -\cos\theta_{\star}\xi^{5} -2\sin\theta_{\star}\xi^{4} +4\cos\theta_{\star}\xi^{3} +6\sin\theta_{\star}\xi^{2} -3\cos\theta_{\star}\xi \right) \right\}, \label{fra4}  \\
\mathcal{Q}_{\mathrm{II}}\left(\theta_{\star};\varphi=0\right) &\equiv \Re\left\{\int_{\xi_{\mathrm{i}}}^{\xi_{\mathrm{f}}}d\xi\,
\frac{e^{\mathrm{i}\left(a\xi+b\right)}}{\left(1+\xi^{2}\right)^{2}}\left( \cos\theta_{\star}\xi^{3} +2\sin\theta_{\star}\xi^{2} -\cos\theta_{\star}\xi \right) \right\}, \label{fra5}  
\end{align} 
with $a\equiv\omega_{\mathrm{g}}r_{\mathrm{eq}}/c$, $b\equiv-\omega_{\mathrm{g}}t_{0}$, and $\xi_{\mathrm{i}}=\cos\theta_{\star}/\left(1+\sin\theta_{\star}\right)$ 
and $\xi_{\mathrm{f}}=0$, and the integrals ${\textstyle {\displaystyle \int}}d\xi\,\xi^{n}e^{\mathrm{i}\left(a\xi+b\right)}/\left(1+\xi^{2}\right)^{m}$ ($m,n$: integers)
being given in terms of the exponential integral $E_{1}\left(z\right)\equiv{\displaystyle \int}_{\!\!\!\! z}^{\infty}\left(e^{-t}/t\right)\mathrm{d}t$ ($\left\vert \arg z\right\vert <\pi$).

\begin{figure}
\centering
\subfloat[]{
		\includegraphics[angle=0, width=7.7cm]{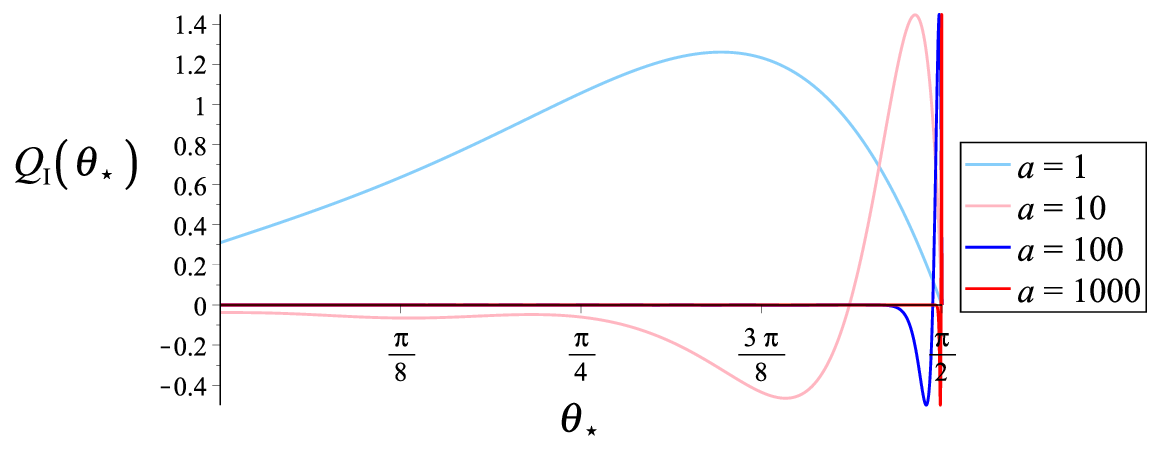}
		\label{fig:subfig01}} 
\hspace{12pt}
\subfloat[]{
		\includegraphics[angle=0, width=7.7cm]{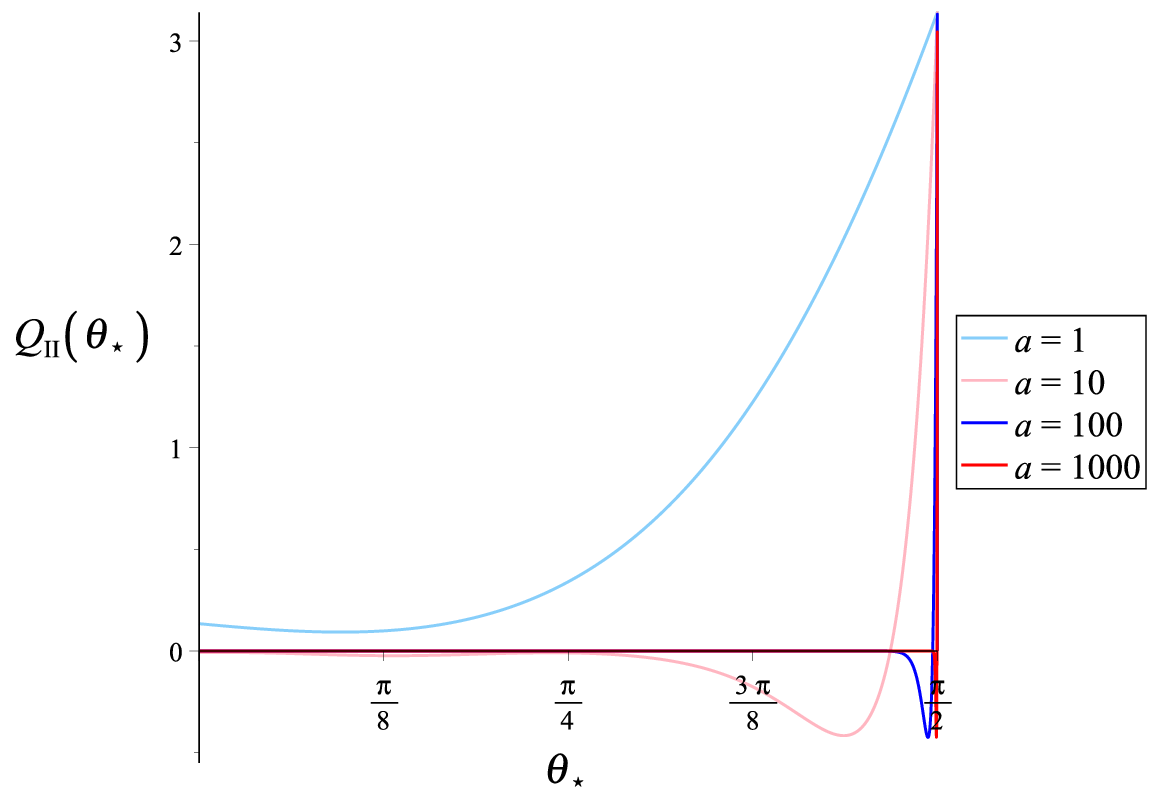}
		\label{fig:subfig02}} % \quad\ \ \ \ \smallskip
\caption{Plots of (a) $\mathcal{Q}_{\mathrm{I}}\left(\theta_{\star};a=1 \right)$, $\mathcal{Q}_{\mathrm{I}}\left(\theta_{\star};a=10 \right)$, $\mathcal{Q}_{\mathrm{I}}\left(\theta_{\star};a=100 \right)$, 
$\mathcal{Q}_{\mathrm{I}}\left(\theta_{\star};a=1000 \right)$ and (b) $\mathcal{Q}_{\mathrm{II}}\left(\theta_{\star};a=1 \right)$, $\mathcal{Q}_{\mathrm{II}}\left(\theta_{\star};a=10 \right)$, $\mathcal{Q}_{\mathrm{II}}\left(\theta_{\star};a=100 \right)$, $\mathcal{Q}_{\mathrm{II}}\left(\theta_{\star};a=1000 \right)$ against $\theta_{\star}$.}
\label{fig_Q}
\end{figure}

Numerical computations of Eqs. (\ref{fra4}) and (\ref{fra5}) at $b=\pi$ provide the maximum values as follows:
\begin{align}
&\mathcal{Q}_{\mathrm{I}}\left(\theta_{\star}\approx62.469^{\,\circ};a=1\right)\approx1.261,\;\; \mathcal{Q}_{\mathrm{I}}\left(\theta_{\star}\approx86.652^{\,\circ};a=10\right)\approx1.446, \notag \\
&\mathcal{Q}_{\mathrm{I}}\left(\theta_{\star}\approx89.664^{\,\circ};a=100\right)\approx1.449,\;\; \mathcal{Q}_{\mathrm{I}}\left(\theta_{\star}\approx89.966^{\,\circ};a=1000\right)\approx1.449, \label{Qimax} \\
&\mathcal{Q}_{\mathrm{II}}\left(\theta_{\star}\approx89.999^{\,\circ};a=1\right)\approx3.142,\;\; \mathcal{Q}_{\mathrm{II}}\left(\theta_{\star}\approx89.999^{\,\circ};a=10\right)\approx3.141, \notag \\
&\mathcal{Q}_{\mathrm{II}}\left(\theta_{\star}\approx89.999^{\,\circ};a=100\right)\approx3.135,\;\; \mathcal{Q}_{\mathrm{II}}\left(\theta_{\star}\approx89.999^{\,\circ};a=1000\right)\approx3.079. \label{Qiimax} 
\end{align}
In Fig. \ref{fig_Q} are plotted $\mathcal{Q}_{\mathrm{I/II}}\left(\theta_{\star};a=1 \right)$, $\mathcal{Q}_{\mathrm{I/II}}\left(\theta_{\star};a=10 \right)$, $\mathcal{Q}_{\mathrm{I/II}}\left(\theta_{\star};a=100 \right)$
and $\mathcal{Q}_{\mathrm{I/II}}\left(\theta_{\star};a=1000 \right)$ against $\theta_{\star}$. 

However, the \textit{dimensionless} factor on the right hand side of Eq. (\ref{fra3}) can be reduced to a simpler, more convenient form by introducing some parameters:
\begin{equation}
\frac{G\mu R_{\mathrm{o}}^{2}\omega_{\mathrm{g}}^{3}}{c^{5}}=\frac{\sqrt{2}q}{p^{5/2}\left(1+q\right)^{2}}, \label{fct}
\end{equation}
where we have substituted $M_{1}=M$, $M_{2}=qM$ into $\mu=M_{1}M_{2}/\left(M_{1}+M_{2}\right)$, $R_{\mathrm{o}}=p\cdot2G\left(M_{1}+M_{2}\right)/c^{2}$ and 
$\omega_{\mathrm{g}}=2\sqrt{G\left(M_{1}+M_{2}\right)/R_{\mathrm{o}}^{3}}$, with $q$ being the mass ratio for the binary system and $p$ being the ratio of the orbital separation 
to the Schwarzschild radius for the binary system. This factor physically means the strain amplitude of our GWs at a distance of the wavelength. It is interesting to note that the factor 
depends only upon the ratios $p$ and $q$; that is, these are the only physical parameters we require to know in order to determine the particular strain amplitude.

\vspace{12pt}
\noindent
\textbf{Example 1: a supermassive BH binary in Mrk 231}

\noindent
A supermassive BH binary candidate in the core of the nearest quasar, Mrk 231 is estimated to have the masses $M_{1}\approx1.5\times10^{8}\, M_{\odot}$, 
$M_{2}\approx4\times10^{6}\, M_{\odot}$ and the orbital separation $R_{\mathrm{o}}=190\times2G\left(M_{1}+M_{2}\right)/c^{2}\approx8.67\times10^{13}\,\mathrm{m}$ \cite{Yan_2015}; 
thus, $q\approx0.03$, $p\approx190$ and the GW frequency $\omega_{\mathrm{g}}=2\sqrt{c^{2}/\left( 2pR_{\mathrm{o}}^{2} \right)}\approx3.55\times10^{-7}\,\mathrm{Hz}$. 
Then from Eq. (\ref{fra3}), the maximum Faraday rotation angles are obtained 
by substituting $q\approx0.03$ and $p\approx190$ into Eq. (\ref{fct}) and combining this with Eqs. (\ref{Qimax}) and (\ref{Qiimax}):
\begin{align} 
\vartheta_{\mathrm{F,I\, max}}&=\vartheta_{\mathrm{F,I}}\left(\theta_{\star}\rightarrow89.664^{\,\circ}\sim89.966^{\,\circ};\varphi=0, a\rightarrow100\sim1000\right) \approx 1.16\times10^{-7}\,\mathrm{rad}, 
\label{framaxi1} \\
\vartheta_{\mathrm{F,II\, max}}&=\vartheta_{\mathrm{F,II}}\left(\theta_{\star}\rightarrow89.999^{\,\circ};\varphi=0, a\rightarrow100\sim1000\right) \approx 2.49\times10^{-7}\,\mathrm{rad}. \label{framaxii1}
\end{align}

\noindent
\textbf{Example 2: a double pulsar system, PSR J0737–3039}

\noindent
PSR J0737-3039 is the first known double pulsar system, consisting of two neutron stars, both being pulsars, having masses $M_{1}\approx1.3381\, M_{\odot}$ and $M_{2}\approx1.2489\, M_{\odot}$,
with its present orbital separation being estimated to be $R_{\mathrm{o}}\approx1.29 \times 10^{9}\,\mathrm{m}\approx 1.7 \times 10^{5}\times2G\left(M_{1}+M_{2}\right)/c^{2}$ \cite{Liu_2021}; 
thus, $q\approx0.9333$, $p\approx1.7 \times 10^{5}$ and the GW frequency $\omega_{\mathrm{g}}=2\sqrt{c^{2}/\left( 2pR_{\mathrm{o}}^{2} \right)}\approx8\times10^{-4}\,\mathrm{Hz}$. 
Then, from Eqs. (\ref{fra3}) and (\ref{Qimax})-(\ref{fct}), the maximum Faraday rotation angles are obtained as
\begin{align} 
\vartheta_{\mathrm{F,I\, max}}&=\vartheta_{\mathrm{F,I}}\left(\theta_{\star}\rightarrow62.469^{\,\circ};\varphi=0, a\rightarrow1\right) \approx 3.74\times10^{-14}\,\mathrm{rad}, 
\label{framaxi2} \\
\vartheta_{\mathrm{F,II\, max}}&=\vartheta_{\mathrm{F,II}}\left(\theta_{\star}\rightarrow89.999^{\,\circ};\varphi=0, a\rightarrow1\right) \approx 9.31\times10^{-14}\,\mathrm{rad}. \label{framaxii2}
\end{align}
However, this double pulsar system evolves very slowly, and the orbit shrinks gradually by GW radiation during the evolution process. In about $8.83 \times 10^{7}$ years (the merger age), 
the orbital separation will reach its minimum around the end of the inspiral phase (or the beginning of the merger phase): 
$R_{\mathrm{o}}\sim2\:\mathrm{NS\:radii}\approx3\times10^{4}\,\mathrm{m}\approx 3.9\times2G\left(M_{1}+M_{2}\right)/c^{2}$ \cite{Liu_2021}; thus, $p\approx3.9$ and the GW frequency 
$\omega_{\mathrm{g}}=2\sqrt{c^{2}/\left( 2pR_{\mathrm{o}}^{2} \right)}\approx7200\,\mathrm{Hz}$. Then, from Eqs. (\ref{fra3}) and (\ref{Qimax})-(\ref{fct}), 
the maximum Faraday rotation angles will be obtained as
\begin{align} 
\vartheta_{\mathrm{F,I\, max}}&=\vartheta_{\mathrm{F,I}}\left(\theta_{\star}\rightarrow62.469^{\,\circ};\varphi=0, a\rightarrow1\right) \approx 0.015\,\mathrm{rad} \approx 0.86^{\,\circ}, \label{framaxi3} \\
\vartheta_{\mathrm{F,II\, max}}&=\vartheta_{\mathrm{F,II}}\left(\theta_{\star}\rightarrow89.999^{\,\circ};\varphi=0, a\rightarrow1\right) \approx 0.037\,\mathrm{rad} \approx 2.12^{\,\circ}. \label{framaxii3}
\end{align}
which are remarkably large, measurable quantities in comparison with those in (\ref{framaxi1})-(\ref{framaxii2}).

\section{Summary and discussion}
\label{concl}
We have investigated the interaction between EMWs and GWs, and its effects on the properties of light, in a situation where light emitted from the neighborhood of a binary meets GWs from the binary 
all along the propagation path towards an observer. The EMW-GW interaction can occur close to the GW source and continues all along the EMW propagation; therefore, the strain amplitude can be initially large 
and decreases inversely as the distance from the source to a point of EMW-GW intersection. As a result, the cumulative effect from the EMW-GW interaction over the long propagation path can be outstanding 
in comparison with the cases, in which EMWs interact with plane GWs, where the strain amplitude is treated as an extremely small constant value far away from the GW source. 
To this end, we have solved Maxwell's equations defined in a spacetime curved due to spherical GWs (i.e., transverse-traceless radially propagating waves). Based on a geometrical-optics analysis, 
our solutions, as given by (\ref{ansatz}), are expressed in terms of first-order modulations of the phase, the wave vector and the polarization vector of EMWs due to the GWs, 
with the exception of the amplitude. In particular, we have focused on the effect of gravitational Faraday rotation (or Skrotskii/Rytov effect), a consequence of the modulation of 
the polarization vectors of EMWs due to the GWs, as described by (\ref{fra1}) and (\ref{fra2}). As applications for this effect, we have calculated the maximum Faraday rotation angles for the two examples: 
(1) a supermassive BH binary in Mrk 231 and (2) a double pulsar system, PSR J0737–3039. Our evaluations of the angle are given by (\ref{framaxi1})-(\ref{framaxii2}),
for both the cases during the inspiral phase; these are much larger than the usual scale of the strain amplitude of GWs reaching the Earth, but still far too small to measure practically. 
However, for the latter case, assuming the system is evolving around the end of the inspiral phase (or the beginning of the merger phase), with the orbital separation $R_{\mathrm{o}}\approx3\times10^{4}\,\mathrm{m}$,
of the same order as the Schwarzschild radius for the binary system, the maximum Faraday rotation angles are estimated to be as large as about $0.86^{\,\circ}$ and $2.12^{\,\circ}$, as given by (\ref{framaxi3}) and (\ref{framaxii3}). This is a fairly interesting result, considering the measurability of the effect from an actual astrophysical event; it will take place practically about $8.83 \times 10^{7}$ years (the merger age) in the future, though. Another noteworthy point is the difference in the Faraday rotation angles for the two polarization vectors $\boldsymbol{\varepsilon}_{\mathrm{I}\left[ 0\right]}$ and 
$\boldsymbol{\varepsilon}_{\mathrm{II}\left[ 0\right]}$. This has been checked out by all the examples above, and implies the anisotropy of the spacetime for light propagation when GWs are present.

In this study, we have assumed a simple theoretical model for a binary with the constant orbital frequency (monochromatic), suitable only for a circular orbit system, non-evolving or extremely slowly evolving. However, 
in order to handle more feasible cases, in which the orbit is quasi-circular (inspiralling) or non-circular, we need to modify our model such that both the orbital frequency and separation change with respect to time 
along the orbit. It will definitely add greater complexity to our analysis, and we leave this discussion to follow-up studies.

\section*{Acknowledgements}
The author appreciates Remo Ruffini for his hospitality at the 18th Italian-Korean Symposium on Relativistic Astrophysics in ICRANet, Pescara, June 19-23, 2023, where part of this work was presented. 
The author was supported by the Basic Science Research Program through the National Research Foundation of Korea (NRF) funded by the Ministry of Education (NRF-2021R1I1A1A01054781).

\appendix

\section{The oblique cylinder}
\label{appA}
The cylindrical surface in Fig. \ref{fig_cyl} is expressed by 
\begin{equation}
\left( x-z\tan \theta _{\star }\right) ^{2}+y^{2}=r_{\mathrm{eq}}^{2}.
\label{cyl}
\end{equation}%
Here the point $\left( x,y,z\right) $ on the surface is written in the spherical polar coordinates $\left( r,\theta ,\phi \right) $ as 
\begin{equation}
\left[ 
\begin{array}{l}
x \\ 
y \\ 
z%
\end{array}%
\right] =\left[ 
\begin{array}{l}
r\sin \theta \cos \phi \\ 
r\sin \theta \sin \phi \\ 
r\cos \theta%
\end{array}%
\right] .  \label{cyl1}
\end{equation}%
Alternatively, however, the surface can be expressed via%
\begin{equation}
\left[ 
\begin{array}{l}
x \\ 
y%
\end{array}%
\right] =\left[ 
\begin{array}{l}
r\tan \theta _{\star }\cos \theta +r_{\mathrm{eq}}\cos \varphi \\ 
r_{\mathrm{eq}}\sin \varphi%
\end{array}%
\right] ,  \label{cyl2}
\end{equation}%
where $\varphi $ is an azimuth specially defined on the circle of radius $r_{\mathrm{eq}}$ (see Fig. \ref{fig_cyl} for comparison of $\varphi $ and $\phi $).
Comparing (\ref{cyl1}) and (\ref{cyl2}), we obtain%
\begin{equation}
r = \frac{r_{\mathrm{eq}}\cos \varphi }{\sin \theta \cos \phi -\tan \theta _{\star}\cos \theta } = \frac{r_{\mathrm{eq}}\sin \varphi }{\sin \theta \sin \phi }.  \label{ratio1}
\end{equation}%

Now, let the angle between the propagation directions of light and GWs be $\gamma $. This can be defined by means of Eqs. (\ref{K0}) and (\ref{k}): 
\begin{equation}
\mathbf{K}_{\left[ 0\right] }\cdot \mathbf{k}=c^{-2}\omega _{\mathrm{e}%
}\omega _{\mathrm{g}}\cos \gamma ,  \label{Kk}
\end{equation}%
where 
\begin{equation}
\cos \gamma \equiv \cos \theta \cos \theta _{\star }+\sin \theta \sin \theta
_{\star }\cos \phi .  \label{gamma}
\end{equation}%
Combining the first equality of Eq. (\ref{ratio1}) and Eq. (\ref{gamma}) together, we obtain
\begin{equation}
r = \frac{r_{\mathrm{eq}}\cos \theta _{\star }\sin \theta _{\star }\cos \varphi }{\cos \theta _{\star }\cos \gamma -\cos \theta }.
\label{ratio2}
\end{equation}

From the triangular region in Fig. \ref{fig_cyl}, one can establish the relation: 
\begin{equation}
r = \frac{r_{\mathrm{eq}}\sin \delta }{\sin \gamma } = \frac{r_{\mathrm{eq}}\sqrt{1-\sin ^{2}\theta _{\star }\cos ^{2}\varphi}\left(1+\xi^{2}\right)}{2\xi},
\label{ratio3}
\end{equation}%
where $\delta $ has been defined via $\cos \delta =-\mathbf{\hat{K}}_{\left[ 0\right] }\cdot \left( \cos \varphi ,\sin \varphi ,0\right) 
=-\sin \theta_{\star }\cos \varphi $, using (\ref{K0}), and $\sin\gamma=2\xi/\left(1+\xi^{2}\right)$ has been substituted, with $\xi\equiv\tan\left(\gamma/2\right)$. 
Then identifying this with Eqs. (\ref{ratio1}) and (\ref{ratio2}), we obtain
\begin{align}
\sin\theta\cos\phi & = {\frac{2\cos^{2}\theta_{\star}\cos\varphi\xi}{\sqrt{1-\sin^{2}\theta_{\star}\cos^{2}\varphi}\left(1+\xi^{2}\right)}
+\frac{\sin\theta_{\star}\left( 1-\xi^{2} \right)}{1+\xi^{2}}},\label{sthcpha} \\
\sin\theta\sin\phi & = {\frac{2\sin\varphi\xi}{\sqrt{1-\sin^{2}\theta_{\star}\cos^{2}\varphi}\left(1+\xi^{2}\right)}},\label{sthspha} \\
\cos\theta & = {\frac{\cos\theta_{\star}\left(1-\xi^{2}\right)}{1+\xi^{2}}
-\frac{2\cos\theta_{\star}\sin\theta_{\star}\cos\varphi\xi}{\sqrt{1-\sin^{2}\theta_{\star}\cos^{2}\varphi}\left(1+\xi^{2}\right)}},\label{ctha} 
\end{align}
where $\cos\gamma=\left(1-\xi^{2}\right)/\left(1+\xi^{2}\right)$ and $\sin\gamma=2\xi/\left(1+\xi^{2}\right)$ have been substituted.

Note that all these quantities are parameterized by $\cos\gamma$ and $\sin\gamma$ while we fix the values of $\theta _{\star}$ and $\varphi$, 
which designate a light beam's direction of propagation and location on the cylindrical surface, respectively.

\bibliographystyle{elsarticle-harv}  
\bibliography{refs_master}

\begin{thebibliography}{26}
\expandafter\ifx\csname natexlab\endcsname\relax\def\natexlab#1{#1}\fi
\providecommand{\url}[1]{\texttt{#1}}
\providecommand{\href}[2]{#2}
\providecommand{\path}[1]{#1}
\providecommand{\DOIprefix}{doi:}
\providecommand{\ArXivprefix}{arXiv:}
\providecommand{\URLprefix}{URL: }
\providecommand{\Pubmedprefix}{pmid:}
\providecommand{\doi}[1]{\href{http://dx.doi.org/#1}{\path{#1}}}
\providecommand{\Pubmed}[1]{\href{pmid:#1}{\path{#1}}}
\providecommand{\bibinfo}[2]{#2}
\ifx\xfnm\relax \def\xfnm[#1]{\unskip,\space#1}\fi
%Type = Article
\bibitem[{Abbott and et~al.(2016)}]{PhysRevLett.116.061102}
\bibinfo{author}{Abbott, B.P.}, \bibinfo{author}{et~al.}
  (\bibinfo{collaboration}{LIGO Scientific Collaboration and Virgo
  Collaboration}), \bibinfo{year}{2016}.
\newblock \bibinfo{title}{Observation of gravitational waves from a binary
  black hole merger}.
\newblock \bibinfo{journal}{Phys. Rev. Lett.} \bibinfo{volume}{116},
  \bibinfo{pages}{061102}.
\newblock \URLprefix
  \url{https://link.aps.org/doi/10.1103/PhysRevLett.116.061102},
  \DOIprefix\doi{10.1103/PhysRevLett.116.061102}.
%Type = Misc
\bibitem[{Amaro-Seoane et~al.(2012)Amaro-Seoane, Aoudia, Babak, Binétruy,
  Berti, Bohé, Caprini, Colpi, Cornish, Danzmann, Dufaux, Gair, Jennrich,
  Jetzer, Klein, Lang, Lobo, Littenberg, McWilliams, Nelemans, Petiteau,
  Porter, Schutz, Sesana, Stebbins, Sumner, Vallisneri, Vitale, Volonteri and
  Ward}]{amaroseoane_2012}
\bibinfo{author}{Amaro-Seoane, P.}, \bibinfo{author}{Aoudia, S.},
  \bibinfo{author}{Babak, S.}, \bibinfo{author}{Binétruy, P.},
  \bibinfo{author}{Berti, E.}, \bibinfo{author}{Bohé, A.},
  \bibinfo{author}{Caprini, C.}, \bibinfo{author}{Colpi, M.},
  \bibinfo{author}{Cornish, N.J.}, \bibinfo{author}{Danzmann, K.},
  \bibinfo{author}{Dufaux, J.F.}, \bibinfo{author}{Gair, J.},
  \bibinfo{author}{Jennrich, O.}, \bibinfo{author}{Jetzer, P.},
  \bibinfo{author}{Klein, A.}, \bibinfo{author}{Lang, R.N.},
  \bibinfo{author}{Lobo, A.}, \bibinfo{author}{Littenberg, T.},
  \bibinfo{author}{McWilliams, S.T.}, \bibinfo{author}{Nelemans, G.},
  \bibinfo{author}{Petiteau, A.}, \bibinfo{author}{Porter, E.K.},
  \bibinfo{author}{Schutz, B.F.}, \bibinfo{author}{Sesana, A.},
  \bibinfo{author}{Stebbins, R.}, \bibinfo{author}{Sumner, T.},
  \bibinfo{author}{Vallisneri, M.}, \bibinfo{author}{Vitale, S.},
  \bibinfo{author}{Volonteri, M.}, \bibinfo{author}{Ward, H.},
  \bibinfo{year}{2012}.
\newblock \bibinfo{title}{elisa: Astrophysics and cosmology in the millihertz
  regime}.
\newblock \URLprefix \url{https://arxiv.org/abs/1201.3621},
  \href{http://arxiv.org/abs/1201.3621}{{\tt arXiv:1201.3621}}.
%Type = Article
\bibitem[{Burgay et~al.(2003)Burgay, D’Amico, Possenti, Manchester, Lyne,
  Joshi, McLaughlin, Kramer, Sarkissian, Camilo, Kalogera, Kim and
  Lorimer}]{Burgay_2003}
\bibinfo{author}{Burgay, M.}, \bibinfo{author}{D’Amico, N.},
  \bibinfo{author}{Possenti, A.}, \bibinfo{author}{Manchester, R.N.},
  \bibinfo{author}{Lyne, A.G.}, \bibinfo{author}{Joshi, B.C.},
  \bibinfo{author}{McLaughlin, M.A.}, \bibinfo{author}{Kramer, M.},
  \bibinfo{author}{Sarkissian, J.M.}, \bibinfo{author}{Camilo, F.},
  \bibinfo{author}{Kalogera, V.}, \bibinfo{author}{Kim, C.},
  \bibinfo{author}{Lorimer, D.R.}, \bibinfo{year}{2003}.
\newblock \bibinfo{title}{An increased estimate of the merger rate of double
  neutron stars from observations of a highly relativistic system}.
\newblock \bibinfo{journal}{Nature} \bibinfo{volume}{426},
  \bibinfo{pages}{531–533}.
\newblock \URLprefix \url{http://dx.doi.org/10.1038/nature02124},
  \DOIprefix\doi{10.1038/nature02124}.
%Type = Article
\bibitem[{Cabral and Lobo(2017)}]{Cabral_2017}
\bibinfo{author}{Cabral, F.}, \bibinfo{author}{Lobo, F.S.N.},
  \bibinfo{year}{2017}.
\newblock \bibinfo{title}{Gravitational waves and electrodynamics: new
  perspectives}.
\newblock \bibinfo{journal}{Eur. Phys. J. C} \bibinfo{volume}{77},
  \bibinfo{pages}{237}.
\newblock \URLprefix \url{https://doi.org/10.1140/epjc/s10052-017-4791-z},
  \DOIprefix\doi{10.1140/epjc/s10052-017-4791-z}.
%Type = Article
\bibitem[{Cruise(1983)}]{Cruise_1983}
\bibinfo{author}{Cruise, A.M.}, \bibinfo{year}{1983}.
\newblock \bibinfo{title}{{An interaction between gravitational and
  electromagnetic waves}}.
\newblock \bibinfo{journal}{Monthly Notices of the Royal Astronomical Society}
  \bibinfo{volume}{204}, \bibinfo{pages}{485--492}.
\newblock \URLprefix
  \url{https://academic.oup.com/mnras/article-pdf/204/2/485/18521706/mnras204-0485.pdf},
  \DOIprefix\doi{10.1093/mnras/204.2.485}.
%Type = Article
\bibitem[{Dewdney et~al.(2009)Dewdney, Hall, Schilizzi and
  Lazio}]{Dewdney_2009}
\bibinfo{author}{Dewdney, P.E.}, \bibinfo{author}{Hall, P.J.},
  \bibinfo{author}{Schilizzi, R.T.}, \bibinfo{author}{Lazio, T.J.L.W.},
  \bibinfo{year}{2009}.
\newblock \bibinfo{title}{The square kilometre array}.
\newblock \bibinfo{journal}{Proceedings of the IEEE} \bibinfo{volume}{97},
  \bibinfo{pages}{1482--1496}.
\newblock \DOIprefix\doi{10.1109/JPROC.2009.2021005}.
%Type = Article
\bibitem[{Dolan(2018)}]{dolan2018geometrical}
\bibinfo{author}{Dolan, S.R.}, \bibinfo{year}{2018}.
\newblock \bibinfo{title}{Geometrical optics for scalar, electromagnetic and
  gravitational waves on curved spacetime}.
\newblock \bibinfo{journal}{International Journal of Modern Physics D}
  \bibinfo{volume}{27}, \bibinfo{pages}{1843010}.
\newblock \URLprefix \url{https://doi.org/10.1142/S0218271818430101},
  \DOIprefix\doi{10.1142/S0218271818430101}.
%Type = Article
\bibitem[{Grote and (forthe LIGO Scientific~Collaboration)(2010)}]{Grote_2010}
\bibinfo{author}{Grote, H.}, \bibinfo{author}{(forthe LIGO
  Scientific~Collaboration)}, \bibinfo{year}{2010}.
\newblock \bibinfo{title}{The geo 600 status}.
\newblock \bibinfo{journal}{Classical and Quantum Gravity}
  \bibinfo{volume}{27}, \bibinfo{pages}{084003}.
\newblock \URLprefix \url{https://dx.doi.org/10.1088/0264-9381/27/8/084003},
  \DOIprefix\doi{10.1088/0264-9381/27/8/084003}.
%Type = Article
\bibitem[{Hacyan(2012)}]{Hacyan_2012}
\bibinfo{author}{Hacyan, S.}, \bibinfo{year}{2012}.
\newblock \bibinfo{title}{Electromagnetic waves and stokes parameters in the
  wake of a gravitational wave}.
\newblock \bibinfo{journal}{General Relativity and Gravitation}
  \bibinfo{volume}{44}, \bibinfo{pages}{2923--2931}.
\newblock \URLprefix \url{https://doi.org/10.1007/s10714-012-1434-4},
  \DOIprefix\doi{10.1007/s10714-012-1434-4}.
%Type = Article
\bibitem[{Hacyan(2016)}]{Hacyan_2016}
\bibinfo{author}{Hacyan, S.}, \bibinfo{year}{2016}.
\newblock \bibinfo{title}{Effects of gravitational waves on the polarization of
  pulsars}.
\newblock \bibinfo{journal}{International Journal of Modern Physics A}
  \bibinfo{volume}{31}, \bibinfo{pages}{1641023}.
\newblock \URLprefix \url{http://dx.doi.org/10.1142/S0217751X16410232},
  \DOIprefix\doi{10.1142/s0217751x16410232}.
%Type = Article
\bibitem[{Halilsoy and Gurtug(2007)}]{Halilsoy_2007}
\bibinfo{author}{Halilsoy, M.}, \bibinfo{author}{Gurtug, O.},
  \bibinfo{year}{2007}.
\newblock \bibinfo{title}{Search for gravitational waves through the
  electromagnetic faraday rotation}.
\newblock \bibinfo{journal}{Phys. Rev. D} \bibinfo{volume}{75},
  \bibinfo{pages}{124021}.
\newblock \URLprefix \url{https://link.aps.org/doi/10.1103/PhysRevD.75.124021},
  \DOIprefix\doi{10.1103/PhysRevD.75.124021}.
%Type = Article
\bibitem[{Hobbs(2013)}]{Hobbs_2013}
\bibinfo{author}{Hobbs, G.}, \bibinfo{year}{2013}.
\newblock \bibinfo{title}{The parkes pulsar timing array}.
\newblock \bibinfo{journal}{Classical and Quantum Gravity}
  \bibinfo{volume}{30}, \bibinfo{pages}{224007}.
\newblock \URLprefix \url{https://dx.doi.org/10.1088/0264-9381/30/22/224007},
  \DOIprefix\doi{10.1088/0264-9381/30/22/224007}.
%Type = Article
\bibitem[{{Hulse} and {Taylor}(1975)}]{Hulse_1975}
\bibinfo{author}{{Hulse}, R.}, \bibinfo{author}{{Taylor}, J.},
  \bibinfo{year}{1975}.
\newblock \bibinfo{title}{{Discovery of a pulsar in a binary system.}}
\newblock \bibinfo{journal}{The Astrophysical Journal Letters}
  \bibinfo{volume}{195}, \bibinfo{pages}{L51--L53}.
\newblock \DOIprefix\doi{10.1086/181708}.
%Type = Misc
\bibitem[{Iyer et~al.(2015)Iyer, Souradeep, Unnikrishnan, Dhurandhar, Raja and
  Sengupta}]{LIGO-India_2011}
\bibinfo{author}{Iyer, B.}, \bibinfo{author}{Souradeep, T.},
  \bibinfo{author}{Unnikrishnan, C.}, \bibinfo{author}{Dhurandhar, S.},
  \bibinfo{author}{Raja, S.}, \bibinfo{author}{Sengupta, A.},
  \bibinfo{year}{2015}.
\newblock \bibinfo{title}{{Hubble Finds That the Nearest Quasar Is Powered by a
  Double Black Hole}}.
\newblock
  \bibinfo{howpublished}{\url{https://dcc.ligo.org/ligo-m1100296/public}}.
%Type = Article
\bibitem[{Kim and Park(2021)}]{Kim_2021}
\bibinfo{author}{Kim, D.H.}, \bibinfo{author}{Park, C.}, \bibinfo{year}{2021}.
\newblock \bibinfo{title}{Detection of gravitational waves by light
  perturbation}.
\newblock \bibinfo{journal}{Eur. Phys. J. C} \bibinfo{volume}{81},
  \bibinfo{pages}{563}.
\newblock \URLprefix \url{https://doi.org/10.1140/epjc/s10052-021-09369-1},
  \DOIprefix\doi{10.1140/epjc/s10052-021-09369-1}.
%Type = Article
\bibitem[{Kramer and Champion(2013)}]{Kramer_2013}
\bibinfo{author}{Kramer, M.}, \bibinfo{author}{Champion, D.J.},
  \bibinfo{year}{2013}.
\newblock \bibinfo{title}{The european pulsar timing array and the large
  european array for pulsars}.
\newblock \bibinfo{journal}{Classical and Quantum Gravity}
  \bibinfo{volume}{30}, \bibinfo{pages}{224009}.
\newblock \URLprefix \url{https://dx.doi.org/10.1088/0264-9381/30/22/224009},
  \DOIprefix\doi{10.1088/0264-9381/30/22/224009}.
%Type = Article
\bibitem[{Liu et~al.(2021)Liu, Yang, Zhang and Rah}]{Liu_2021}
\bibinfo{author}{Liu, P.}, \bibinfo{author}{Yang, Y.Y.},
  \bibinfo{author}{Zhang, J.W.}, \bibinfo{author}{Rah, M.},
  \bibinfo{year}{2021}.
\newblock \bibinfo{title}{Simulation of the orbit and spin period evolution of
  the double pulsars psr j0737-3039 from their birth to coalescence induced by
  gravitational wave radiation}.
\newblock \bibinfo{journal}{Research in Astronomy and Astrophysics}
  \bibinfo{volume}{21}, \bibinfo{pages}{104}.
\newblock \URLprefix \url{https://dx.doi.org/10.1088/1674-4527/21/4/104},
  \DOIprefix\doi{10.1088/1674-4527/21/4/104}.
%Type = Article
\bibitem[{Lyne et~al.(2004)Lyne, Burgay, Kramer, Possenti, Manchester, Camilo,
  McLaughlin, Lorimer, D’Amico, Joshi, Reynolds and Freire}]{Lyne_2004}
\bibinfo{author}{Lyne, A.G.}, \bibinfo{author}{Burgay, M.},
  \bibinfo{author}{Kramer, M.}, \bibinfo{author}{Possenti, A.},
  \bibinfo{author}{Manchester, R.}, \bibinfo{author}{Camilo, F.},
  \bibinfo{author}{McLaughlin, M.A.}, \bibinfo{author}{Lorimer, D.R.},
  \bibinfo{author}{D’Amico, N.}, \bibinfo{author}{Joshi, B.C.},
  \bibinfo{author}{Reynolds, J.}, \bibinfo{author}{Freire, P.C.C.},
  \bibinfo{year}{2004}.
\newblock \bibinfo{title}{A double-pulsar system: A rare laboratory for
  relativistic gravity and plasma physics}.
\newblock \bibinfo{journal}{Science} \bibinfo{volume}{303},
  \bibinfo{pages}{1153–1157}.
\newblock \URLprefix \url{http://dx.doi.org/10.1126/science.1094645},
  \DOIprefix\doi{10.1126/science.1094645}.
%Type = Book
\bibitem[{Maggiore(2007)}]{Maggiore_2007}
\bibinfo{author}{Maggiore, M.}, \bibinfo{year}{2007}.
\newblock \bibinfo{title}{{Gravitational Waves: Volume 1: Theory and
  Experiments}}.
\newblock \bibinfo{publisher}{Oxford University Press}, \bibinfo{address}{New
  York}.
\newblock \URLprefix
  \url{https://doi.org/10.1093/acprof:oso/9780198570745.001.0001}.
%Type = Article
\bibitem[{Manchester(2013)}]{Manchester_2013}
\bibinfo{author}{Manchester, R.N.}, \bibinfo{year}{2013}.
\newblock \bibinfo{title}{The international pulsar timing array}.
\newblock \bibinfo{journal}{Classical and Quantum Gravity}
  \bibinfo{volume}{30}, \bibinfo{pages}{224010}.
\newblock \URLprefix \url{https://dx.doi.org/10.1088/0264-9381/30/22/224010},
  \DOIprefix\doi{10.1088/0264-9381/30/22/224010}.
%Type = Misc
\bibitem[{{NASA Hubble Mission Team}(2015)}]{Hubble2015}
\bibinfo{author}{{NASA Hubble Mission Team}}, \bibinfo{year}{2015}.
\newblock \bibinfo{title}{{Hubble Finds That the Nearest Quasar Is Powered by a
  Double Black Hole}}.
\newblock
  \bibinfo{howpublished}{\url{https://science.nasa.gov/centers-and-facilities/goddard/hubble-finds-that-the-nearest-quasar-is-powered-by-a-double-black-hole/}}.
%Type = Article
\bibitem[{Park and Kim(2021)}]{Park_2021}
\bibinfo{author}{Park, C.}, \bibinfo{author}{Kim, D.H.}, \bibinfo{year}{2021}.
\newblock \bibinfo{title}{Observation of gravitational waves by light
  polarization}.
\newblock \bibinfo{journal}{Eur. Phys. J. C} \bibinfo{volume}{81},
  \bibinfo{pages}{95}.
\newblock \URLprefix \url{https://doi.org/10.1140/epjc/s10052-021-08893-4},
  \DOIprefix\doi{10.1140/epjc/s10052-021-08893-4}.
%Type = Book
\bibitem[{Poisson and Will(2014)}]{poisson2014gravity}
\bibinfo{author}{Poisson, E.}, \bibinfo{author}{Will, C.},
  \bibinfo{year}{2014}.
\newblock \bibinfo{title}{Gravity: Newtonian, Post-Newtonian, Relativistic}.
\newblock \bibinfo{publisher}{Cambridge University Press},
  \bibinfo{address}{New York}.
\newblock \URLprefix \url{https://books.google.co.kr/books?id=PZ5cAwAAQBAJ}.
%Type = Article
\bibitem[{Somiya and (for~the KAGRA~Collaboration)(2012)}]{Somiya_2012}
\bibinfo{author}{Somiya, K.}, \bibinfo{author}{(for~the KAGRA~Collaboration)},
  \bibinfo{year}{2012}.
\newblock \bibinfo{title}{Detector configuration of kagra–the japanese
  cryogenic gravitational-wave detector}.
\newblock \bibinfo{journal}{Classical and Quantum Gravity}
  \bibinfo{volume}{29}, \bibinfo{pages}{124007}.
\newblock \URLprefix \url{https://dx.doi.org/10.1088/0264-9381/29/12/124007},
  \DOIprefix\doi{10.1088/0264-9381/29/12/124007}.
%Type = Book
\bibitem[{Thorne and Blandford(2017)}]{thorne2017modern}
\bibinfo{author}{Thorne, K.}, \bibinfo{author}{Blandford, R.},
  \bibinfo{year}{2017}.
\newblock \bibinfo{title}{Modern Classical Physics: Optics, Fluids, Plasmas,
  Elasticity, Relativity, and Statistical Physics}.
\newblock \bibinfo{publisher}{Princeton University Press},
  \bibinfo{address}{New Jersey}.
\newblock \URLprefix \url{https://books.google.co.kr/books?id=U1S6BQAAQBAJ}.
%Type = Article
\bibitem[{Yan et~al.(2015)Yan, Lu, Dai and Yu}]{Yan_2015}
\bibinfo{author}{Yan, C.S.}, \bibinfo{author}{Lu, Y.}, \bibinfo{author}{Dai,
  X.}, \bibinfo{author}{Yu, Q.}, \bibinfo{year}{2015}.
\newblock \bibinfo{title}{A probable milli-parsec supermassive binary black
  hole in the nearest quasar mrk 231}.
\newblock \bibinfo{journal}{The Astrophysical Journal} \bibinfo{volume}{809},
  \bibinfo{pages}{117}.
\newblock \URLprefix \url{https://dx.doi.org/10.1088/0004-637X/809/2/117},
  \DOIprefix\doi{10.1088/0004-637X/809/2/117}.

\end{thebibliography}



%% For example:

%% \bibitem[Aladro et al.(2015)]{Aladro15} Aladro, R., Martín, S., Riquelme, D., et al. 2015, \aas, 579, A101

%%\end{thebibliography}

\end{document}